\documentclass[aps,pra,twocolumn,showpacs,superscriptaddress]{revtex4}

\usepackage{graphicx}
\usepackage{amsmath}
\usepackage{amssymb}

\begin{document}

\newcommand{\Enl}{\ensuremath{E_{nl}}}
\newcommand{\En}{\ensuremath{E_{n}}}
\newcommand{\Eres}{\ensuremath{E_\text{res}}}
\newcommand{\Bres}{\ensuremath{B_\text{res}}}
\newcommand{\Epres}{\ensuremath{E'_\text{res}}}
\newcommand{\Bel}{\ensuremath{\beta_{El}}}
\newcommand{\Rbef}{\ensuremath{^{85}\text{Rb}}}
\newcommand{\Rbes}{\ensuremath{^{87}\text{Rb}}}
\newcommand{\TF}{\ensuremath{T_{\text{F}}}}
\newcommand{\Prem}{\ensuremath{P_{\text{rem}}}}
\newcommand{\Ptr}{\ensuremath{P_{\text{tr}}}}
\newcommand{\Pcool}{\ensuremath{P_{\text{cool}}}}
\newcommand{\widthE}{\ensuremath{\Gamma_{E}}}
\newcommand{\Eeff}{\ensuremath{E_{\text{eff}}}}
\newcommand{\Zs}{\ensuremath{Z_s}}
\newcommand{\Eavg}{\ensuremath{\langle E_{\text{tot}} \rangle}}
\newcommand{\EavgCM}{\ensuremath{\langle E_{\text{CM}} \rangle}}
\newcommand{\dE}{\ensuremath{\Delta E}}
\newcommand{\Ecut}{\ensuremath{E}_\text{cut}}
\newcommand{\aBG}{\ensuremath{a_{\text{bg}}}}
\newcommand{\BigArctan}{
  \arctan{
    \left(
      \frac
      {a\left( \Enl , B \right) \hbar \omega}
      {2\Losc\Enl\sqrt{e}}
    \right)
  }
}
\newcommand{\derfrac}[2]{\genfrac{}{}{}{}{\text{d}#1}{\text{d}#2}}
\newcommand{\pderfrac}[2]{\genfrac{}{}{}{}{\partial#1}{\partial#2}}
\newcommand{\td}{\ensuremath{\text{d}}}
\newcommand{\half}{\ensuremath{\frac{1}{2}}}
\newcommand{\swave}{\ensuremath{s}-wave}
\newcommand{\pwave}{\ensuremath{p}-wave}
\newcommand{\dwave}{\ensuremath{d}-wave}
\newcommand{\kB}{\ensuremath{k_{\text{\tiny{B}}}}}
\newcommand{\etal}{\emph{et~al.}}
\newcommand{\schrodinger}{Schr\"{o}dinger}
\newcommand{\Hz}{\text{Hz}}
\newcommand{\KHz}{\text{KHz}}
\newcommand{\MHz}{\text{MHz}}
\newcommand{\Gauss}{\text{G}}
\newcommand{\nutrap}{\nu_{\text{trap}}}
\newcommand{\Losc}{L_{\text{osc}}}
\newcommand{\Lau}{L_{\text{au}}}
\newcommand{\Eosc}{\ensuremath{E_{\text{osc}}}}

\def\bra#1{\mathinner{\langle{#1}|}}
\def\ket#1{\mathinner{|{#1}\rangle}}
\def\braket#1{\mathinner{\langle{#1}\rangle}}
\def\Bra#1{\left<#1\right|}
\def\Ket#1{\left|#1\right>}

\title{Feshbach Resonance Cooling of Trapped Atom Pairs}

\author{Josh~W.~Dunn}
\affiliation{Department of Physics and JILA, University of
Colorado, Boulder, Colorado 80309-0440}
\author{D.~Blume}
\affiliation{Department of Physics, Washington State University,
Pullman, Washington 99164-2814}
\author{Bogdan~Borca}
\affiliation{Department of Physics and JILA, University of
Colorado, Boulder, Colorado 80309-0440}
\author{B.~E.~Granger}
\affiliation{Department of Physics, Santa Clara University, Santa
  Clara, CA 95053}
\author{Chris~H.~Greene}
\affiliation{Department of Physics and JILA, University of
Colorado, Boulder, Colorado 80309-0440}

\date{\today}

\begin{abstract}
Spectroscopic studies of few-body systems at ultracold temperatures
provide valuable information that often cannot be extracted in a hot
environment.  Considering a pair of atoms, we propose a cooling
mechanism that makes use of a scattering Feshbach
resonance. Application of a series of time-dependent magnetic field
ramps results in the situation in which either zero, one, or two atoms
remain trapped.  If two atoms remain in the trap after the field ramps
are completed, then they have been cooled.  Application of the
proposed cooling mechanism to optical traps or lattices is considered.
\end{abstract}


\maketitle

A Feshbach resonance~\cite{feshbach92a,stwalley76a,tiesinga93a} occurs
for two atoms when their collision energy becomes degenerate with a
vbound state in a closed collision channel, producing brief transitions
into and out of this state.  In recent years, these resonances have
been used extensively to control the interaction strength in dilute
atomic gases~\cite{inouye98a,courteille98a,roberts98a}.  Here, we
utilize some of the unique characteristics of a Feshbach resonance to
develop a cooling mechanism that is applicable to two
externally-confined atoms.  Application of a series of field ramps
(i.e., cooling cycles) leads to a cooled atom pair (an atom pair with
reduced internal energy), provided that the atom pair remains trapped.

In the following, we first develop the basic mechanism of the Feshbach
resonance cooling process.  The feasibility and effectiveness of the
proposed scheme are then illustrated through an application to a
realistic system of two atoms in a trap.  Finally, possible
applications to optical traps are discussed.

The concept of Feshbach resonance cooling grows out of the observation
that the quantum-mechanical energy levels of two atoms in a harmonic
trap shift by an energy corresponding to approximately two trap
quanta, as a control parameter is swept in one direction across the
resonance.  Throughout this article, we refer to this control
parameter as the magnetic field $B$ used to manipulate the atom-atom
scattering length $a$ in the vicinity of a pole.  In other contexts,
the shift of the energy levels could be introduced by varying the
detuning of an off-resonant dressing laser, or by varying an electric
field strength. The ideas presented here in terms of the control
parameter $B$ can be straightforwardly extended to those other
contexts.

The Schr\"odinger equation for two interacting identical mass $m$
atoms under spherical harmonic confinement with trapping frequency
$\nu$ decouples into two equations: one involving the three relative
coordinates of the pair, and another involving the three
center-of-mass (CM) coordinates~\cite{busch98a,blume02a}.  We consider
the Schr\"odinger equation in the relative coordinate for two trapped
atoms interacting through a central potential and assume for the time
being that the center of mass coordinate is translationally
cold. Accounting for an applied external magnetic field $B$ through a
$B$-dependent quantum defect $\Bel(B)$, the energies $\Enl(B)$
associated with the relative motion of an atom pair are given
by~\cite{blume02a}
\begin{equation}
  \Enl(B)
  =
  \left(
    2 n - 2 \Bel(B) + l + 3/2
  \right)
  \hbar\omega,
  \label{eq:DefectEnergy}
\end{equation}
where $\omega=2\pi\nu$.  Here, the quantum defect $\Bel(B)$ depends
strongly on the relative orbital angular momentum $l$ of the pair,
while it depends only weakly on the radial oscillator quantum number
$n$.  The dependence of $\Bel(B)$ on the energy is weak on the scale
of an oscillator quantum, i.e., $\left| \text{d} \Bel(B) / \text{d}
\Enl \right| \ll 1 / \hbar \omega$.

As will become clear later, the quantum defect for one relative
partial wave $l$ for an atom pair, e.g., the \swave, \pwave, or
\dwave, must rise by unity across the energy range $\kB \Delta T$ of
interest, and across the accessible range of the control parameter,
$\Delta B$.  In fact, this variation of $\Bel(B)$ by unity corresponds
to the Feshbach resonance, which causes the scattering phaseshift to
rise by $\pi$.  A simple closed-form expression exists for
\Bel~\cite{blume02a,bolda02a}, which simplifies at energies higher
than a few trap quanta to $\Bel(B) \approx \BigArctan$ where $a
(\Enl,B)$ is the energy- and field-dependent scattering length and
$\Losc = \sqrt{\hbar / \left( \mu \omega \right)}$ with $\mu = m/2$
denotes the characteristic oscillator length.

\begin{figure}
  \begin{center}
    \includegraphics[width=2.8in]{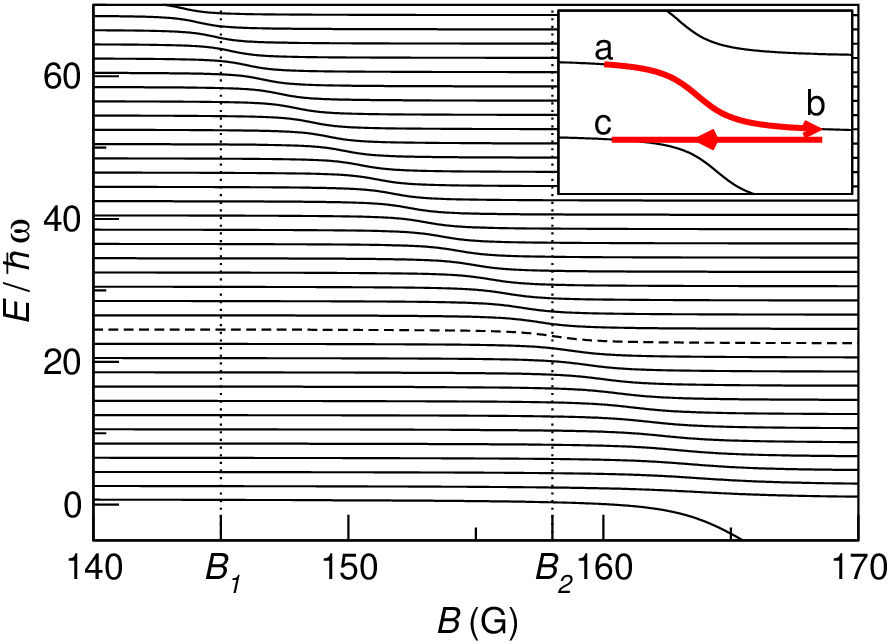}
  \end{center}
  \caption{(Color online) Energy levels $\En$ for the relative
    coordinate of a harmonically trapped $\Rbef$ atom pair near the
    $B_0 \approx 155.2$\,G Feshbach resonance, as a function of the
    magnetic field $B$.  A rather large trapping frequency of
    $\nu=1$\,MHz is used in order to clarify the field dependence of
    the energy levels.  Cooling is performed by ramping the magnetic
    field $B$ slowly from $B_1$ to $B_2$ and then quickly back to
    $B_1$.  A more realistic ramp would likely encompass more level
    curves (i.e., cover a larger field range).  The state which
    undergoes a shift for $B=B_2$ (which we will label as $n=Q$ later)
    is indicated by a dashed line.  Ideal cooling is described
    diagramatically in the inset (same axes), where population
    transfer from point a to point b occurs during the slow field ramp
    and from b to c during the fast ramp.  See the text for details.
    \label{fig:Levels}
  }
\end{figure}

In this paper, we focus on an \swave\ resonance, though this formalism
can be readily extended to higher partial wave resonances.  When an
\swave\ Feshbach resonance occurs, the limiting low-energy scattering
phaseshift is proportional to the wavenumber $k = (2
\mu E/ \hbar^2)^{1/2}$.  Omitting the subscript $l$, the $E$- and
$B$-dependent scattering length is then given by
\begin{equation}
  a(\En,B)
  =
  \aBG
  +
  \frac{\Gamma_E\sqrt{\hbar^2/(8\mu\En)}}{\En+(B-\Bres)\Epres(B)},
  \label{eq:ScattLength}
\end{equation}
where $\aBG$ is the background scattering length.  At the magnetic
field strength $\Bres$ of the resonance a zero-energy bound state
occurs.  The resonance width in energy $\Gamma_E$ is related to the
width in the control parameter $\Delta$ by $\Gamma_E =
2k\aBG\Epres(B)\Delta$, where $\Epres$ denotes the rate at which the
resonance energy $\Eres$ varies with the control
parameter~\cite{mies00a}.  Figure~\ref{fig:Levels} illustrates the
characteristic \swave\ energy levels $E_n$ appropriate for the
relative motion of two atoms in a spherical harmonic oscillator trap,
as functions of the applied magnetic field $B$ for a magnetic Feshbach
resonance in $^{85}$Rb$(2,-2) + ^{85}$Rb$(2,-2)$.  In
Fig.~\ref{fig:Levels}, the parameters adopted are $\Bres =
155.2\,\Gauss$, $\Epres = -3.5\,\MHz/\Gauss$, $\Gamma_B = 10\,\Gauss$,
and $\aBG=-380\,a_0$, where $a_0$ is the Bohr radius.

Feshbach resonance cooling entails ramping the magnetic field through
the region where the energy levels shift by $\approx 2\hbar\omega$.
Figure~\ref{fig:Levels} denotes the initial $B$-field by $B_1$.  For
an atom pair taken from a source of atoms with temperature $T$, the
probability of the pair being in the energy eigenstate $\Ket{n(B_1)}$,
and hence, of having the energy $\En(B_1)$, is given by the
corresponding Boltzmann factor.  By ramping the magnetic field from
$B_1$ to $B_2$ (also shown in the inset of Fig.~\ref{fig:Levels}),
sufficiently slowly to be adiabatic, the population of each energy
level remains unchanged, while the energy level itself is reduced
compared to its value at $B_1$, $\En(B_2) \approx \En(B_1) - 2 \hbar
\omega$; the atom pair has lost energy.  To further reduce the energy
of the two atoms, another field ramp has to be applied.  In order to
do this, it is necessary to return the magnetic field to $B_1$ without
adding the energy back that has been just removed.  A fast,
nonadiabatic change of the magnetic field from $B_2$ to $B_1$ ensures
this, since this change simply projects the eigenstate $\Ket{n(B_2)}$
onto the eigenstate $\Ket{n(B_1)}$.  Note that this proposed energy
reduction (cooling) comes at the expense of an energy gain of a single
level (or a few levels, see below for a detailed discussion).

To model the effects of the magnetic field ramps, we have developed a
two-channel Feshbach resonance model, based on the single-channel
model described in Ref.~\cite{borca03a}.  Both of these models
describe a two-atom Feshbach resonance for a harmonic trap, and
utilize a zero-range potential to describe the interaction between the
two atoms.  The two-channel model has the advantage of allowing for a
field-dependent resonance state.  In the two channels, the \swave\
radial solutions for the relative coordinate $r$ of the atom pair
satisfy the equations
\begin{gather}
  \left(-\frac{\hbar^2}{2\mu}\derfrac{^2}{r^2} +
  \frac{1}{2}\mu\omega r^2\right)u_1(r)
  =
  E u_1(r)\\
  \left(-\frac{\hbar^2}{2\mu}\derfrac{^2}{r^2} +
  \frac{1}{2}\mu\omega r^2\right)u_2(r)
  =
  \left(E - \varepsilon\right) u_2(r),
\end{gather}
where $\varepsilon$ is the energy shift of the second channel from the
first channel.  The zero-range potential imposes a boundary condition
at the origin, which is parameterized as
\begin{equation}
  \derfrac{}{r}\begin{pmatrix}u_1(r)\\u_2(r)\end{pmatrix}_{r=0}
  =
  \begin{pmatrix}-1/a_1&\beta\\\beta&-1/a_2\end{pmatrix}
  \begin{pmatrix}u_1(r)\\u_2(r)\end{pmatrix}_{r=0}.
\end{equation}
A quantum-defect-theory treatment, similar to Ref.~\cite{borca03a},
can then be applied.  The scattering length predicted by this model
(when $\omega \rightarrow 0$) is
\begin{equation}
  a(E,B)
  =
  \left(
    \frac{1}{a_1}
    +
    \frac
    {\left|\beta\right|^2}
    {\sqrt{2\mu\varepsilon(B)/\hbar^2 - 2\mu E/\hbar^2}-1/a_2}
  \right)^{-1},
\end{equation}
which can be compared to the measured scattering length to determine
the values of the parameters $a_1$, $a_2$, and $\beta$.  The
parameters also affect the magnetic-field dependence of the adiabatic
energy states, and their adjustment is able to provide satisfactory
agreement with experimental data in the regions of interest to us.
E.g., for \Rbef, we find $a_1 = -435\,a_0$, $a_2 = 1.485\,a_0$, and
$\beta = 0.0011618\,a_0^{-1}$.  Simulations can then be performed by
specifying an initial state of the system and numerically propagating
the \schrodinger\ equation.

The simulations reveal the effect of the adiabatic and nonadiabatic
field ramps: Assume first that the atom pair is in a pure state at
$B=B_1$. As expected, the adiabatic field ramp ($B_1$ to $B_2$ in
Fig.~\ref{fig:Levels}) decreases the energy of the atom pair
irrespective of the initial eigenstate chosen. A nonadiabatic ramp
($B_2$ to $B_1$ in Fig.~\ref{fig:Levels}) causes a state at $B_2$ that
is not degenerate with the resonance state to project onto a state at
$B_1$, with approximately the same energy as the initial state at
$B_2$. However, if the pair is initially in the state at $B_2$ that is
degenerate with the resonance state, the fast ramp results in a strong
projection onto the resonance state.  In this case, the atom pair
gains energy since the resonance state at $B_1$ has a higher energy
than the initial state at $B_2$.

We now generalize our scheme to the more practical situation of a
mixed initial state, and show how removal of hot atoms leads to an
intriguing cooling scheme.  For a mixed state, the occupation
probability of a pair level with energy $\En$ in the relative motion
is determined in terms of a Boltzmann factor by
$e^{-\En/\tau}/Z(\tau)$ with $\tau=\kB T$, where $\kB$ is Boltzmann's
constant, $Z(\tau)$ is the (relative) partition function, and $T$ is
the temperature of the source of the two atoms.  Based on the results
of field ramps for pure states discussed above, we see that
application of a cooling cycle (slow ramp from $B_1$ to $B_2$ plus
fast ramp back to $B_1$) for a mixed state will do two things: (a)
decrease by $2\hbar\omega$ the energy of the population in states
which undergo a full energy shift between $B_1$ and $B_2$, and (b)
increase the energy of the population in the state that is degenerate
with the resonance (i.e., undergoing an energy shift) at $B_2$ by
moving it to the state (or states) degenerate with the resonance at
$B_1$.  We will denote this state from which the heated population
originates as $n=Q$.

\begin{figure}
  \begin{center}
    \includegraphics[width=2.6in]{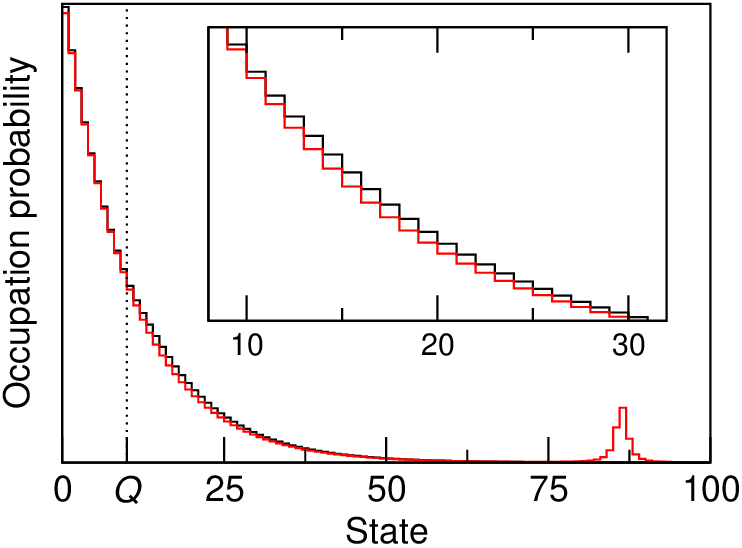}
  \end{center}
  \caption{(Color online) Illustration of the effect of a single
   Feshbach resonance cooling cycle for $T=1\,$mK and $\nu=1\,$MHz.
   The black (red) line represents the population distribution before
   (after) application of one slow and one fast magnetic field ramp.
   The state $Q$ (here $Q=10$) is indicated.  Population from $n=Q$
   and nearby states is moved to higher states with $n \approx 85$.
   Population initially in a state with $n>Q$ is moved to the
   next-lowest state (see the inset close-up).
   \label{fig:Diagram}
  }
\end{figure}

Figure~\ref{fig:Diagram} illustrates the effect of a single cooling
cycle on a mixed state, using the same Feshbach resonance as shown in
Fig.~\ref{fig:Levels}.  The black line represents an initial \swave\
probability distribution for the states associated with the relative
coordinate of an atom pair in a harmonic trap with $\nu=1\,$MHz and
source temperature $T=1\,$mK.  The red line represents the same
probability distribution after application of a slow and a fast
magnetic field ramp, for $Q=10$.  Application of one cooling cycle
moves the population of the state $Q$ to states with much higher
energy, here $n \approx 85$, evidenced by the spike in
Fig.~\ref{fig:Diagram}.  At the same time, the field ramps move the
population of each state with $n>Q$ to the next-lowest state, which
has $\approx 2\hbar\omega$ less energy.  Our numerical simulations
indicate that, on average, the net energy of the system is increased
after application of one cooling cycle.  The few cases where the pair
gains a large amount of energy overcomes the many cases where the pair
looses a small amount of energy.  This behavior is expected, since
Ketterle and Pritchard~\cite{ketterle92a} in 1992 pointed out the
impossibility of creating a cooling scheme relying solely on
time-dependent potentials.

An estimate of the efficiency of cooling can be made by assuming that
the population of the state $Q$ is removed from the trap, while the
population of all states with $n>Q$ are moved to the next-lowest
state, that is, to states with $n-1$.  This assumes that the range of
the field ramps is such that the heated fraction ($n=Q$) ends up at an
energy corresponding to negligible thermal population (this is the
case in Fig.~\ref{fig:Diagram}), and that all population above a
specified energy can be removed.  If we approximate the level energies at
$B=B_1$ by $\En(B_1) \approx 2 n \hbar \omega$, the probability to
remove an atom pair during a cooling cycle is
\begin{equation}
  \Prem (Q,\tau)
  =
  \frac{e^{ -2 \hbar \omega Q / \tau }}{Z(\tau)}.
  \label{eq:Prem}
\end{equation}
The average energy decrease in a cooling cycle is due to the energy of
the $n \approx Q$ population removed from the trap, plus the energy
loss for states $n>Q$:
\begin{multline}
  \dE (Q,\tau)
  =
  \left(
    2 \hbar \omega Q + \EavgCM
  \right)
  \frac{e^{-2 \hbar \omega Q / \tau}}{Z(\tau)}
  \\
  + \sum_{n=Q+1}^{\infty} 2 \hbar \omega
  \frac{e^{-2 \hbar \omega n / \tau}}{Z(\tau)}.
  \label{eq:DeltaE1}
\end{multline}
Noting that $\EavgCM=3\tau$ (since $\Eavg=3\tau$ for a single atom in
a harmonic trap), and with $\sum_{n=Q+1}^{\infty} e^{-2\hbar\omega
n/\tau}\approx e^{-2\hbar\omega Q/\tau}\tau/2\hbar\omega,$
Eq.~(\ref{eq:DeltaE1}) becomes
\begin{equation}
  \Delta E (Q,\tau)
  =
  \frac{e^{-2 \hbar \omega Q / \tau}}{Z(\tau)}
  \left(2 \hbar \omega Q + 4\tau \right).
  \label{eq:DeltaE2}
\end{equation}
The energy efficiency $\Eeff$, defined as the amount of energy removed
per atom removed, is then given by
\begin{equation}
  \Eeff (Q,\tau)
  =
  2\hbar\omega Q + 4\tau.
  \label{eq:Eeff}
\end{equation}
Since $Q$ determines the efficiency of the cooling process, it is
referred to as the cooling parameter.  Results from our numerical
model indicate that Eq.~(\ref{eq:Eeff}) provides a good estimate of
the efficiency.

The time scale for one cooling cycle is determined by the speed of the
adiabatic field ramp.  This speed in turn is determined by the
strength of the coupling between the resonance state and the trap
states.  The smaller the coupling for an avoided crossing, the slower
is the field ramp required to maintain adiabaticity.
The coupling between the resonance
state and the trap states is related to the resonance width parameter
$\Gamma_{E}$, which can be used in a Landau-Zener estimate of the
transition probability~\cite{mies00a},
\begin{equation}
  \Ptr
  \cong
  \exp
  \left(
    -
    \frac{2}{ \left| \text{d}B/\text{d}t \right| }
    \frac{\omega \widthE}{\left| \text{d}E/\text{d}B \right|}
  \right).
  \label{eq:Ptr}
\end{equation}

\begin{figure}
  \begin{center}
    \includegraphics[width=2.6in]{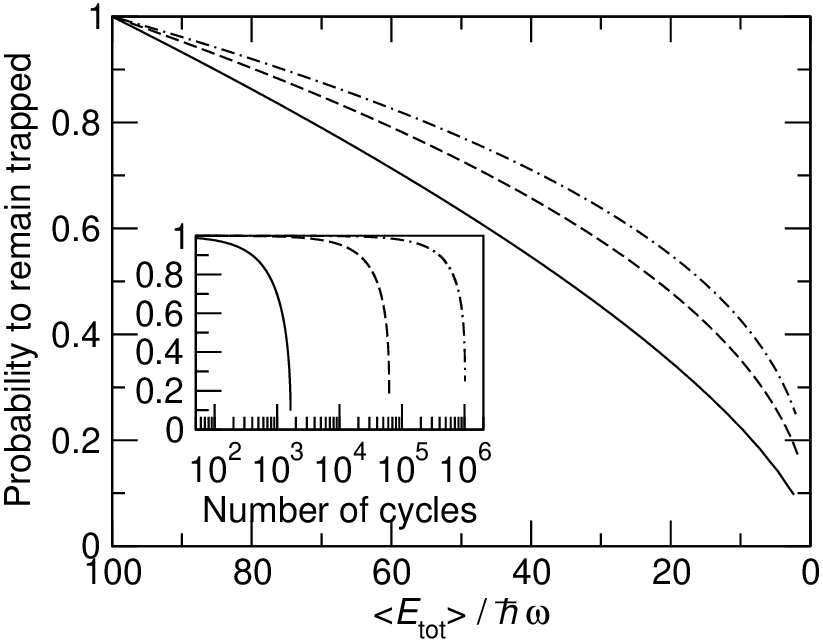}
  \end{center}
  \caption{Probability that a pair of atoms remains trapped vs. the
    average total kinetic energy of the two atoms in oscillator units
    (note that $\kB T=\Eavg/6$ for two harmonically trapped atoms).
    Three different cooling parameters are used: $2\hbar\omega Q =
    5\tau$ (solid line), $9\tau$ (dashed line), and $12\tau$
    (dot-dashed line).  It is assumed that rethermalization occurs
    between cooling cycles (see text).  Inset: probability to remain
    trapped vs. the number of cooling cycles for the same three
    cooling parameters.
    \label{fig:Cooling}
  }
\end{figure}

Motivated by the possibility of experimentally trapping a small,
deterministic number of atoms~\cite{frese00a}, we now explore the
experimental feasibility of our cooling scheme.  A Feshbach resonance
cooling experiment involves a sequence of cooling cycles.  As
discussed above, a single experiment could result in a heated atom
pair, which in turn would be lost from the trap.  To end up with a
cooled atom pair, multiple ramp cycles are required.  To see the
effect of cooling, Eqs.~(\ref{eq:Prem}) and~(\ref{eq:DeltaE1}) can be
iterated.  For a variety of cooling efficiency parameters $Q$,
Fig.~\ref{fig:Cooling} shows the probability for an atom pair to
remain trapped vs. the average total kinetic energy (the energy of
both the relative and the CM degrees of freedom) of the two atoms in
oscillator units.  Included in this calculation is the probability
that the atom pair is in an \swave\ state to begin with, because the
field ramp has no effect on other partial waves.  We assume that
rethermalization occurs between cooling cycles, which could be ensured
by, for example, introducing a slight anharmonicity into the trapping
potential.


To be more specific, we consider a crossed-beam optical dipole
trap~\cite{barrett01a} which offers a good blend of large trap
frequency (for a large \swave\ fraction), isotropy, and anharmonicity
(for rethermalization between the relative and CM degrees of freedom).
Assuming the dipole trap has an average frequency of $\nu=10\,$kHz and
contains two atoms taken from a source with temperature $T=8\,\mu$K
($\Eavg/\hbar\omega = 100$), we see from Fig.~\ref{fig:Cooling}, solid
line, that a temperature of $~0.16\,\mu$K ($\Eavg/\hbar\omega = 2$,
both atoms in the ground state) could be reached $10\%$ of the time by
performing less than 20,000 cooling cycles.  For a range in magnetic
field for the ramps of $\Delta B \approx 1\,$G, and using
Eq.~(\ref{eq:Ptr}) with $\Ptr = 0.1$, we see that such a series of
field ramps could take place in under $1\,$s.  A pertubative
calculation accounting for the trap anharmonicity (see
Ref.~\cite{bolda04a} for details of a similar treatment) indicates
that rethermalization between the relative and CM degrees of freedom
should occur on a time scale comparable to a single ramp time for a
crossed-beam dipole trap.  This will ensure that the relative \swave\
distribution will rethermalize with each ramp and that the cooling of
the relative coordinate will also cool the CM coordinate (both of
which we have assumed up to this point).

Another possibility would be to apply our cooling scheme to atom pairs
trapped in an optical lattice.  In this case, field ramps could be
performed on the lattice ensemble of atom pairs, with a certain
percentage of sites resulting in cooled pairs, while other sites will
have either zero atoms or one (uncooled) atom.  It may also be
possible to prepare the optical lattice by some other means to have a
high probability of exactly double occupancy at each lattice cite (see
for example Ref.~\cite{rabl03a}).  From such an initial state, a
Feshbach resonance cooling scheme could be used to efficiently cool
atom pairs to low-lying trap states.

In summary, we have proposed a novel cooling scheme that makes use of
Feshbach resonances.  We have also shown that such a cooling scheme
may be experimentally feasible, although it may not be competitive
with proven cooling methods like evaporation.  At the same time, plenty of
room for exploration remains.  Extension to atom clouds might also be
possible.

This work was supported in part by the National Science Foundation.
We thank M. Holland, C. Wieman, E. Cornell, P. Julienne, D. Jin, and
M. Baertschy for helpful discussions.


\begin{thebibliography}{16}
\expandafter\ifx\csname natexlab\endcsname\relax\def\natexlab#1{#1}\fi
\expandafter\ifx\csname bibnamefont\endcsname\relax
  \def\bibnamefont#1{#1}\fi
\expandafter\ifx\csname bibfnamefont\endcsname\relax
  \def\bibfnamefont#1{#1}\fi
\expandafter\ifx\csname citenamefont\endcsname\relax
  \def\citenamefont#1{#1}\fi
\expandafter\ifx\csname url\endcsname\relax
  \def\url#1{\texttt{#1}}\fi
\expandafter\ifx\csname urlprefix\endcsname\relax\def\urlprefix{URL }\fi
\providecommand{\bibinfo}[2]{#2}
\providecommand{\eprint}[2][]{\url{#2}}

\bibitem[{\citenamefont{Feshbach}(1992)}]{feshbach92a}
\bibinfo{author}{\bibfnamefont{H.}~\bibnamefont{Feshbach}},
  \emph{\bibinfo{title}{Theoretical Nuclear Physics}}
  (\bibinfo{publisher}{Wiley}, \bibinfo{address}{New York, NY},
  \bibinfo{year}{1992}).

\bibitem[{\citenamefont{Stwalley}(1976)}]{stwalley76a}
\bibinfo{author}{\bibfnamefont{W.~C.} \bibnamefont{Stwalley}},
  \bibinfo{journal}{Phys. Rev. Lett.} \textbf{\bibinfo{volume}{37}},
  \bibinfo{pages}{1628} (\bibinfo{year}{1976}).

\bibitem[{\citenamefont{Tiesinga et~al.}(1993)\citenamefont{Tiesinga, Verhaar,
  and Stoof}}]{tiesinga93a}
\bibinfo{author}{\bibfnamefont{E.}~\bibnamefont{Tiesinga}},
  \bibinfo{author}{\bibfnamefont{B.~J.} \bibnamefont{Verhaar}},
  \bibnamefont{and} \bibinfo{author}{\bibfnamefont{H.~T.~C.}
  \bibnamefont{Stoof}}, \bibinfo{journal}{Phys. Rev. A}
  \textbf{\bibinfo{volume}{47}}, \bibinfo{pages}{4114} (\bibinfo{year}{1993}).

\bibitem[{\citenamefont{Inouye et~al.}(1998)\citenamefont{Inouye, Andrews,
  Stenger, Miesner, Stamper-Kurn, and Ketterle}}]{inouye98a}
\bibinfo{author}{\bibfnamefont{S.}~\bibnamefont{Inouye}},
  \bibinfo{author}{\bibfnamefont{M.~R.} \bibnamefont{Andrews}},
  \bibinfo{author}{\bibfnamefont{J.}~\bibnamefont{Stenger}},
  \bibinfo{author}{\bibfnamefont{H.-J.} \bibnamefont{Miesner}},
  \bibinfo{author}{\bibfnamefont{D.~M.} \bibnamefont{Stamper-Kurn}},
  \bibnamefont{and} \bibinfo{author}{\bibfnamefont{W.}~\bibnamefont{Ketterle}},
  \bibinfo{journal}{Nature} \textbf{\bibinfo{volume}{392}},
  \bibinfo{pages}{151} (\bibinfo{year}{1998}).

\bibitem[{\citenamefont{Courteille et~al.}(1998)\citenamefont{Courteille,
  Freeland, Heinzen, van Abeelen, and Verhaar}}]{courteille98a}
\bibinfo{author}{\bibfnamefont{P.}~\bibnamefont{Courteille}},
  \bibinfo{author}{\bibfnamefont{R.~S.} \bibnamefont{Freeland}},
  \bibinfo{author}{\bibfnamefont{D.~J.} \bibnamefont{Heinzen}},
  \bibinfo{author}{\bibfnamefont{F.~A.} \bibnamefont{van Abeelen}},
  \bibnamefont{and} \bibinfo{author}{\bibfnamefont{B.~J.}
  \bibnamefont{Verhaar}}, \bibinfo{journal}{Phys. Rev. Lett.}
  \textbf{\bibinfo{volume}{81}}, \bibinfo{pages}{69} (\bibinfo{year}{1998}).

\bibitem[{\citenamefont{Roberts et~al.}(1998)\citenamefont{Roberts, Claussen,
  {Burke,~Jr.}, Greene, Cornell, and Wieman}}]{roberts98a}
\bibinfo{author}{\bibfnamefont{J.~L.} \bibnamefont{Roberts}},
  \bibinfo{author}{\bibfnamefont{N.~R.} \bibnamefont{Claussen}},
  \bibinfo{author}{\bibfnamefont{J.~P.} \bibnamefont{{Burke,~Jr.}}},
  \bibinfo{author}{\bibfnamefont{C.~H.} \bibnamefont{Greene}},
  \bibinfo{author}{\bibfnamefont{E.~A.} \bibnamefont{Cornell}},
  \bibnamefont{and} \bibinfo{author}{\bibfnamefont{C.~E.}
  \bibnamefont{Wieman}}, \bibinfo{journal}{Phys. Rev. Lett.}
  \textbf{\bibinfo{volume}{81}}, \bibinfo{pages}{5109} (\bibinfo{year}{1998}).

\bibitem[{\citenamefont{Busch et~al.}(1998)\citenamefont{Busch, Englert,
  Rza\c{z}\.{e}wski, and Wilkins}}]{busch98a}
\bibinfo{author}{\bibfnamefont{T.}~\bibnamefont{Busch}},
  \bibinfo{author}{\bibfnamefont{B.-G.} \bibnamefont{Englert}},
  \bibinfo{author}{\bibfnamefont{K.}~\bibnamefont{Rza\c{z}\.{e}wski}},
  \bibnamefont{and} \bibinfo{author}{\bibfnamefont{M.}~\bibnamefont{Wilkins}},
  \bibinfo{journal}{Found. Phys.} \textbf{\bibinfo{volume}{28}},
  \bibinfo{pages}{549} (\bibinfo{year}{1998}).

\bibitem[{\citenamefont{Blume and Greene}(2002)}]{blume02a}
\bibinfo{author}{\bibfnamefont{D.}~\bibnamefont{Blume}} \bibnamefont{and}
  \bibinfo{author}{\bibfnamefont{C.~H.} \bibnamefont{Greene}},
  \bibinfo{journal}{Phys. Rev. A} \textbf{\bibinfo{volume}{65}},
  \bibinfo{pages}{043613} (\bibinfo{year}{2002}).

\bibitem[{\citenamefont{Bolda et~al.}(2002)\citenamefont{Bolda, Tiesinga, and
  Julienne}}]{bolda02a}
\bibinfo{author}{\bibfnamefont{E.~L.} \bibnamefont{Bolda}},
  \bibinfo{author}{\bibfnamefont{E.}~\bibnamefont{Tiesinga}}, \bibnamefont{and}
  \bibinfo{author}{\bibfnamefont{P.~S.} \bibnamefont{Julienne}},
  \bibinfo{journal}{Phys. Rev. A} \textbf{\bibinfo{volume}{66}},
  \bibinfo{pages}{013403} (\bibinfo{year}{2002}).

\bibitem[{\citenamefont{Mies et~al.}(2000)\citenamefont{Mies, Tiesinga, and
  Julienne}}]{mies00a}
\bibinfo{author}{\bibfnamefont{F.~H.} \bibnamefont{Mies}},
  \bibinfo{author}{\bibfnamefont{E.}~\bibnamefont{Tiesinga}}, \bibnamefont{and}
  \bibinfo{author}{\bibfnamefont{P.~S.} \bibnamefont{Julienne}},
  \bibinfo{journal}{Phys. Rev. A} \textbf{\bibinfo{volume}{61}},
  \bibinfo{pages}{022721} (\bibinfo{year}{2000}).

\bibitem[{\citenamefont{Borca et~al.}(2003)\citenamefont{Borca, Blume, and
  Greene}}]{borca03a}
\bibinfo{author}{\bibfnamefont{B.}~\bibnamefont{Borca}},
  \bibinfo{author}{\bibfnamefont{D.}~\bibnamefont{Blume}}, \bibnamefont{and}
  \bibinfo{author}{\bibfnamefont{C.~H.} \bibnamefont{Greene}},
  \bibinfo{journal}{New J. Phys.} \textbf{\bibinfo{volume}{5}},
  \bibinfo{pages}{111.1} (\bibinfo{year}{2003}).

\bibitem[{\citenamefont{Ketterle and Pritchard}(1992)}]{ketterle92a}
\bibinfo{author}{\bibfnamefont{W.}~\bibnamefont{Ketterle}} \bibnamefont{and}
  \bibinfo{author}{\bibfnamefont{D.~E.} \bibnamefont{Pritchard}},
  \bibinfo{journal}{Phys. Rev. A} \textbf{\bibinfo{volume}{46}},
  \bibinfo{pages}{4051} (\bibinfo{year}{1992}).

\bibitem[{\citenamefont{Frese et~al.}(2000)\citenamefont{Frese, Ueberholz,
  Kuhr, Alt, Schrader, Gomer, and Meschede}}]{frese00a}
\bibinfo{author}{\bibfnamefont{D.}~\bibnamefont{Frese}},
  \bibinfo{author}{\bibfnamefont{B.}~\bibnamefont{Ueberholz}},
  \bibinfo{author}{\bibfnamefont{S.}~\bibnamefont{Kuhr}},
  \bibinfo{author}{\bibfnamefont{W.}~\bibnamefont{Alt}},
  \bibinfo{author}{\bibfnamefont{D.}~\bibnamefont{Schrader}},
  \bibinfo{author}{\bibfnamefont{V.}~\bibnamefont{Gomer}}, \bibnamefont{and}
  \bibinfo{author}{\bibfnamefont{D.}~\bibnamefont{Meschede}},
  \bibinfo{journal}{Phys. Rev. Lett.} \textbf{\bibinfo{volume}{85}},
  \bibinfo{pages}{3777} (\bibinfo{year}{2000}).

\bibitem[{\citenamefont{Barrett et~al.}(2001)\citenamefont{Barrett, Sauer, and
  Chapman}}]{barrett01a}
\bibinfo{author}{\bibfnamefont{M.}~\bibnamefont{Barrett}},
  \bibinfo{author}{\bibfnamefont{J.}~\bibnamefont{Sauer}}, \bibnamefont{and}
  \bibinfo{author}{\bibfnamefont{M.~S.} \bibnamefont{Chapman}},
  \bibinfo{journal}{Phys. Rev. Lett.} \textbf{\bibinfo{volume}{87}},
  \bibinfo{pages}{010404} (\bibinfo{year}{2001}).

\bibitem[{\citenamefont{Bolda et~al.}(2004)\citenamefont{Bolda, Tiesinga, and
  Julienne}}]{bolda04a}
\bibinfo{author}{\bibfnamefont{E.~L.} \bibnamefont{Bolda}},
  \bibinfo{author}{\bibfnamefont{E.}~\bibnamefont{Tiesinga}}, \bibnamefont{and}
  \bibinfo{author}{\bibfnamefont{P.~S.} \bibnamefont{Julienne}},
  \bibinfo{journal}{to be published}  (\bibinfo{year}{2004}).

\bibitem[{\citenamefont{Rabl et~al.}(2003)\citenamefont{Rabl, Daley, Fedichev,
  Cirac, and Zoller}}]{rabl03a}
\bibinfo{author}{\bibfnamefont{P.}~\bibnamefont{Rabl}},
  \bibinfo{author}{\bibfnamefont{A.~J.} \bibnamefont{Daley}},
  \bibinfo{author}{\bibfnamefont{P.~O.} \bibnamefont{Fedichev}},
  \bibinfo{author}{\bibfnamefont{J.~I.} \bibnamefont{Cirac}}, \bibnamefont{and}
  \bibinfo{author}{\bibfnamefont{P.}~\bibnamefont{Zoller}},
  \bibinfo{journal}{Phys. Rev. Lett.} \textbf{\bibinfo{volume}{91}},
  \bibinfo{pages}{110403} (\bibinfo{year}{2003}).

\end{thebibliography}


\end{document}